\documentclass[aps,prb,superscriptaddress,twocolumn,showpacs]{revtex4}

\usepackage{amsmath,amssymb} 
\usepackage{graphicx}

\newfont{\multilead}{multilead}
\newcommand{\threelead}{\hbox{\multilead A}}
\newcommand{\fourlead}{\hbox{\multilead B}}

\begin{document}

\title{Effects of Exchange Symmetry on Full Counting Statistics}

\author{F. Hassler} 
\affiliation{Theoretische Physik,
  ETH Zurich, CH-8093 Zurich, Switzerland}

\author{G.B. Lesovik}
\affiliation{L.D.\ Landau Institute for Theoretical Physics RAS,
   117940 Moscow, Russia}

\author{G. Blatter}
\affiliation{Theoretische Physik,
  ETH Zurich, CH-8093 Zurich, Switzerland}

\date{\today}

\begin{abstract} 
  We study the full counting statistics for the transmission of two identical
  particles with positive or negative symmetry under exchange for the
  situation where the scattering depends on energy.  We find that, besides the
  expected sensitivity of the noise and higher cumulants, the exchange
  symmetry has a huge effect on the average transmitted charge; for equal-spin
  exchange-correlated electrons, the average transmitted charge can be orders
  of magnitude larger than the corresponding value for independent electrons.
  A similar, although smaller, effect is found in a four-lead geometry even
  for energy-independent scattering.
\end{abstract}


\maketitle

\section{Introduction}

Full counting statistics provides the ultimate information on the
statistics of charge transport in mesoscopic systems.  The first study
of this type \cite{levitov:93} already provided a surprising result, a
(quantum \cite{levitov:92,binomial_s}) binomial distribution of charges,
telling that the flux of incoming electrons is regular (a consequence
of exchange correlations) and the scattering events are independent.
This result applies to the situation where the scatterer is characterized
by an energy-independent transmission and for a constant applied voltage.
The modification due to adiabatic pumping \cite{muzykantskii:03} leaves
the binomial result largely unchanged.  Recently, much interest has
concentrated on single particle sources feeding devices with individual
electrons;\cite{lee:95,levitov:96,ivanov:97,lebedev:05,keeling:06} such
single particle excitations are generated by (unit-flux) voltage pulses.
Again, it turns out that the full counting analysis remains binomial when the
system is driven by such voltage pulses.\cite{ivanov:97} These findings make
explicit that exchange correlations are absent in the (energy-independent)
scattering process. In this letter, we demonstrate that exchange effects
manifest themselves when single particle sources inject electrons into
devices with energy dependent scattering.

\begin{figure}
  \centering 
  \includegraphics{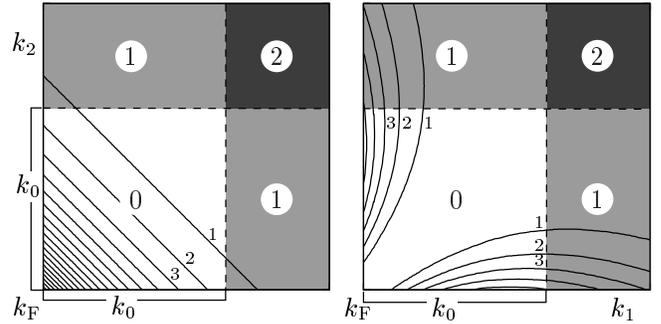}
  \caption{Contour plots for the two-particle wave functions
   $|\Psi_{\text{in},\pm}(k_1,k_2)|^2$ with vanishing separation $\delta x$
   between the particles. The three processes with 0, 1, and 2 particles
   transmitted draw their weight $P_n$, $n=0,1,2$, from the (shaded)
   regions labelled with 0, 1, and 2 (shown is the case for the quantum
   point contact, with a Lorentzian wave function Eq.~(\ref{eq:wf_lorentz}).)}
\label{fig:redist}
\end{figure}

The analysis of a setup with energy-dependent scattering and driven with
a \textit{constant} voltage has unveiled sensitivity to exchange in the
noise but not in the average current.\cite{buttiker:92} Furthermore,
exchange effects in the noise were found in the statistics of two
particles escaping from a dot at short times \cite{lesovik:habil} and
for two electrons traversing a reflectionless symmetric beam splitter
\cite{burkard:00}. Here, we show that applying voltage \textit{pulses} in
the incoming lead and injecting individual particles produces non-linear
transport due to exchange effects; i.e., a twice larger voltage pulse does
not double the transmitted charge and hence the average current is sensitive
to the exchange symmetry.  Note that voltage pulses applied \textit{across}
the device are not expected to generate single particle excitations.

The discussion of full counting statistics in the presence of an
energy-dependent scattering is difficult to conduct in a second-quantized
formalism. On the other hand, the recent insight regarding the correspondence
between fidelity and full counting statistics \cite{lesovik:06:c} has enabled
a first-quantized approach based on a wave-packet formalism. Here, we make
use of this new approach in our investigation of exchange effects on the full
counting statistics of charge transport and find interesting new results:
while it is expected that the noise as well as higher-order correlators
will be sensitive to the exchange symmetry, we find the astounding result
that exchange can hugely enhance (or suppress) the \emph{average charge}
already.  The effect is a consequence of the symmetry-induced reshuffling
of weight in the momentum distribution of the two-particle wave function,
see Fig.~\ref{fig:redist}, combined with the energy dependence of the
scattering matrix: the anti-symmetry under exchange moves weight away from
the (Pauli-blocked) diagonal, which typically leads to an enhancement in the
scattering channel transmitting one particle, at the expense of the channel
where both particles are reflected. In a multi-channel/multi-lead setup
the effect appears already for energy-independent transmission amplitudes.

\section{Single Lead}

\subsection{First Quantized Picture}

\begin{figure}[b]
  \centering 
  \includegraphics{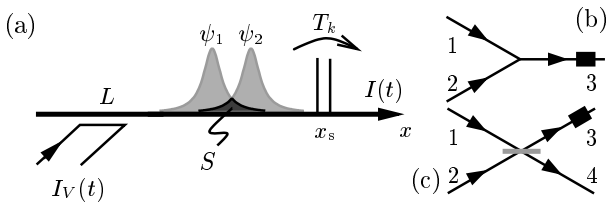} 
  \caption{(a) Single channel quantum wire driven inductively ($L$) with a
   nearby current $I_V(t)$.  Two Lorentzian voltage pulses $V(t)
   = L \dot{I}_V/c $ drive the wave packets $\psi_1$ and $\psi_2$
   with overlap $S$.  The scatterer at $x=x_\text{s}$ gives rise to a
   momentum-dependent transmission probability $T_k$.  (b) Three lead fork
   geometry with incoming wave packets in leads 1 and 2 and the counter
   (black box) in lead 3.  (c) Reflectionless four-lead beam splitter;
   the incoming wave packets in leads 1 and 2 undergo a reflectionless
   transmission into leads 3 and 4.}
\label{fig:setup}
\end{figure}

We start with the discussion of a quantum wire with one conducting
channel, cf.\ Fig.~\ref{fig:setup}(a). The task is to find the full
counting statistics of the charge transport through a scatterer with an
energy-dependent transmission amplitude located at $x=x_\text{s}$. Using
the wave-packet formalism of Ref.\ \onlinecite{lesovik:06:c}, we calculate
the generating function for the full counting statistics for two incoming
particles, the simplest case exhibiting the effect of exchange symmetry. The
incoming particles are described by two wave packets of the form
\begin{equation}\label{eq:wave_packet}
  \psi_{\text{in},m}(x;t) = \int_0^\infty
  \frac{dk}{2\pi} f_m(k) e^{i k (x - v_\text{F} t)}
\end{equation}
with the Fourier components $f_1(k)$ and $f_2(k)$. The normalization of the
wave packets in momentum space reads $\int (dk/2\pi) | f_m(k) |^2 = 1$.  At
low temperatures the interesting physics takes place near the Fermi points and
we can use a linearized spectrum $\epsilon = v_{\rm\scriptscriptstyle F} |k|$
with the Fermi velocity $v_{\rm\scriptscriptstyle F}$, the momentum $\hbar k$
and the energy $\hbar \epsilon$. Traversing the scatterer, the wave packet,
Eq.~(\ref{eq:wave_packet}), acquires a reflected and transmitted term
\begin{multline}\label{eq:wf_out}
  \psi^\sigma_{\text{out},m} (x,t) = \int_0^\infty 
  \frac{dk}{2\pi} f_m(k) e^{- i k v_\text{F} t} 
  \Bigl[ r_k e^{-ikx} \Theta(x_\text{s}-x) \\
  + e^{i \sigma \lambda/2} \tau_k e^{ikx} \Theta(x-x_\text{s}) \Bigr],
\end{multline}
where $\tau_k$ ($r_k$) is the transmission (reflection) amplitude and we have
introduced a counting field $\exp(i\sigma \lambda/2)$ in the transmitted
part; the sign $\sigma = \pm 1$ differentiates between the forward- and
back-propagating wavefunctions required in the wave-packet formalism of full
counting statistics, cf.\ Ref.\ \onlinecite{lesovik:06:c}.  The properly
(anti)sym\-metrized two-particle wave functions assume the form $\Psi_{{\rm
x},\pm}(x_1,x_2;t)\propto\psi_{{\rm x},1}(x_1;t)\psi_{{\rm x},2}(x_2;t)\pm
(x_1\leftrightarrow x_2)$ with ${\rm x} = {\rm in},~{\rm out}$ (see,
Fig.~\ref{fig:setup}(c); throughout the paper, the subscript `$\pm$'
refers to the exchange symmetry, e.g., for two electrons this corresponds
to the triplet or singlet states). The transport statistics is described by
the generating function $\chi_\pm (\lambda) = \int dx_1 dx_2\, \Psi_{{\rm
out},\pm}^{-1\,*} \Psi_{{\rm out},\pm}^{+1}$ and we obtain the result
\begin{multline}\label{eq:fcs_2wp}
  \chi_\pm (\lambda) = \frac{
    \bigl[ 1 + (e^{i \lambda} - 1 ) \langle 1 | T | 1 \rangle \bigr] 
    \bigl[1 + (e^{i \lambda} - 1 )  \langle 2 | T | 2 \rangle \bigr] } 
    {1 \pm |S|^2} \\
  \pm \frac{ 
    \bigl[ S + (e^{i \lambda} - 1 ) \langle 1 | T | 2 \rangle \bigr] 
    \bigl[ S^* + (e^{i \lambda} - 1 ) \langle 2 | T | 1 \rangle\bigr] } 
    {1 \pm |S|^2}
\end{multline}
with the matrix elements $\langle n | T | m \rangle = \int (dk/2\pi) f_n^*(k)
T_k$ $f_m(k)$ involving the transmission probability $T_k = |\tau_k|^2$ and
the overlap integral $S = \int (dk/2\pi) f_1^*(k) f_2(k)$.\cite{lesovik:06:c}
The Fourier transform $P_n = \int (d\lambda/2\pi) \chi(\lambda) e^{-i \lambda
n}$ yields the probability $P_n$ for transmitting $n$ particles. 

\begin{figure}[t] 
  \centering
  \includegraphics{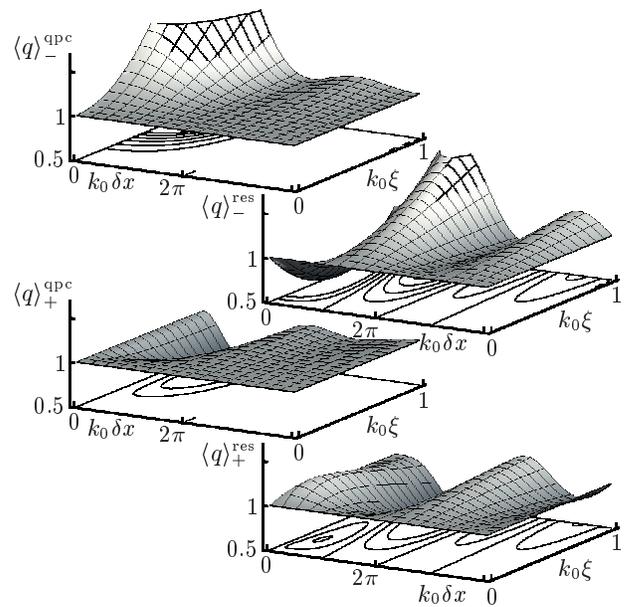}
  \caption{Normalized average charge $q= Q/2e\langle T \rangle$ transmitted
   through a transmission step (left) and a transmission resonance
   (right). The step/resonance is placed at a distance $k_0$ away from the
   Fermi momentum $k_{\rm\scriptscriptstyle F}$, $\xi$ is the width of the
   wave packets in real space, and $\delta x$ denotes the separation between
   the wave packets. The top (bottom) figures show the average transmitted
   charge for particles with negative (positive) exchange symmetry. Note
   the huge enhancement in the average transmitted charge in the limit
   $\delta x \to 0$ and $k_0\xi \geq 1$ for the case of negative exchange.}
   \label{fig:charge}
\end{figure}

For \emph{energy-independent} transmission amplitudes $\tau_k \equiv \tau$,
the exchange term in Eq.~(\ref{eq:fcs_2wp}) cancels with the denominator
and the counting statistics does not dependent on the exchange symmetry of
the particles. This property is particular to the one-dimensional wire; in
a multi-lead setup, this cancellation does not occur and interference terms
can be observed even for energy-independent scattering amplitudes, see below.

On the contrary, for an \emph{energy-dependent} transmission, the
exchange term has a dramatic effect on the statistical properties of the
transferred charge.  To keep the discussion simple, we consider the case of
two incident wave packets with the same momentum distribution separated in
position. Given the initial separation $\delta x$, the Fourier components
of the wave packets satisfy $f_2(k) = f_1(k) e^{- i k \delta x}$ and thus,
$\langle 1 | T | 1 \rangle = \langle 2 | T | 2 \rangle \equiv \langle T
\rangle$ $= \int (dk/2\pi) T_k |f_1(k)|^2$. The overlap integral $S=\int
(dk/2\pi) |f_1(k)|^2 \exp (i k \delta x)$ is the Fourier transform of the
distribution in momentum space. The transmission probabilities $P_{n,\pm}$
are given by ($\sum_n P_{n,\pm}=1$)
\begin{align}\label{eq:prob}
  P_{0,\pm} &= \frac{ ( 1 - \langle T \rangle )^2 
    \pm | S - \langle 1 | T | 2 \rangle |^2 }
    {1 \pm |S|^2}, \nonumber\\
  P_{1,\pm} &= 2 \frac{\langle T \rangle ( 1 - \langle T \rangle) 
    \pm \bigr[\text{Re} ( \langle 1 | T | 2 \rangle S^*) 
    - | \langle 1 | T | 2 \rangle |^2 \bigr]}
    {1 \pm |S|^2}, \nonumber\\
  P_{2,\pm} &= \frac{ \langle T \rangle ^2 
    \pm | \langle 1 | T | 2 \rangle |^2 } 
    {1 \pm |S|^2}; 
\end{align}
they depend on the exchange symmetry provided that $\langle 1 | T | 2 \rangle
\ne S \langle T \rangle$.  These transmission probabilities are easily
converted into cumulants of transmitted charge $Q = \int dt I(t)$.
\cite{gustavsson:06} For two particles incident on the scatterer, the first
two cumulants read ($e$ denotes the charge of the particles)
\begin{align}\label{eq:cum}
  \langle Q/e \rangle_\pm 
  &= P_{1,\pm} + 2 P_{2,\pm} \nonumber \\
  \langle\langle (Q/e)^2 \rangle\rangle_\pm
  &= P_{1,\pm} ( 1- P_{1,\pm} ) + 4 P_{2,\pm} P_{0,\pm} 
\end{align}
and we find both depending on the exchange symmetry.  While interference terms
in the noise $\langle\langle Q^2 \rangle\rangle$ due to the exchange symmetry
of electrons are expected and have been found before,
\cite{buttiker:92,lesovik:habil,burkard:00} here, we find that already the
transmitted charge $\langle Q \rangle$ is sensitive to the exchange symmetry.

For a quantitative analysis of the effect, we need to specify the shape $f_1$
of the wave packets as well as the energy dependence $T_k$ of the transmission
probability.  Rather than postulating trivial Gaussian wave packets, here, we
consider a more realistic situation where the wave packets are deliberately
created by voltage pulses.  The creation of such single-particle excitations
requires voltage pulses of unit flux\cite{levitov:96} $\Phi_0 = hc/e$ and with
Lorentzian shape,\cite{keeling:06} $V_{t_1}(t)=-(2v_{\rm\scriptscriptstyle
F} \xi \Phi_0/c)/[v_{\rm\scriptscriptstyle F}^2( t-t_1)^2 + \xi^2]$, where we
conveniently express the pulse duration $\xi/v_{\rm\scriptscriptstyle F}$
through the length parameter $\xi$.  Such a voltage pulse generates a wave
packet with amplitude ($x_1 = v_{\rm\scriptscriptstyle F}t_1$)
\begin{equation}\label{eq:lorentz}
  f_1(k) = \sqrt{4\pi\xi} e^{-\xi (k-k_{\rm\scriptscriptstyle F})-i k x_1}
  \Theta(k-k_{\rm\scriptscriptstyle F}),
\end{equation}
and a Lorentzian shape in real space, 
\begin{equation}\label{eq:wf_lorentz}
    |\psi_1|^2 =  \frac{\xi/\pi}{(x-x_1-v_{\rm\scriptscriptstyle F}
     t)^2+\xi^2},
\end{equation}
cf.\ Appendix~\ref{sec:equiv}.

For two wave packets separated by $\delta x$ we obtain an overlap integral
$S = e^{-i k_{\rm\scriptscriptstyle F} \delta x}/ (1 + i \delta x/2 \xi)$.

Next, we concentrate on the scatterer where we consider two generic types:
(i) for a sharp \emph{transmission resonance}, e.g., due to a double
barrier, the energy dependent transmission probability assumes the form
$T_k^\text{res} = \alpha/[1 + \beta^2 (k - k_{\rm\scriptscriptstyle F} -
k_0)^2]$ where $\alpha \leq 1$ is the amplitude of the resonance and $k_0>0$
is its position relative to the Fermi wave vector $k_{\rm\scriptscriptstyle
F}$. The width $\beta^{-1}$ of the resonance has to be much smaller
than the width $\xi^{-1}$ of the wave packet in $k$-space, $\beta^{-1}
\ll \xi^{-1}$.  The transmission probability $\langle T^\text{res}
\rangle$ for a single wave packet with amplitude $f_1(k)$ assumes the
form $\langle T^\text{res} \rangle \approx (2 \pi \alpha \xi /\beta) e^{-2
\xi k_0}$ for $\beta^{-1} \ll k_0$, i.e., a resonance away from the Fermi
level. While a small parameter $k_0\xi$ produces a large overall signal,
it does suppress the effect of exchange symmetry since the transmission is
already saturated, hence, below we will be interested in intermediate and
large values of $k_0\xi$. (ii) For a (sharp, $\beta^{-1} \ll \xi^{-1}$)
\emph{transmission step}, e.g., due to a quantum point contact, we use
the Kemble function \cite{landau:3} $T_k^\text{qpc}= \alpha/[1+e^{-\beta
(k-k_{\rm \scriptscriptstyle F} -k_0)}]$ producing the averaged transmission
probability $\langle T^\text{qpc}\rangle\approx \alpha e^{-2\xi k_0}$;
note that the small prefactor $\xi/\beta$ is absent in this case.

For a sharp resonance, the exchange term assumes the simple form $\langle
1 | T^\text{res} | 2 \rangle\approx e^{-i (k_{\rm\scriptscriptstyle F}+k_0) \delta x}
\langle T^\text{res} \rangle$ and its product with the overlap integral
$S^*$ results in an expression $\propto \exp (-i k_0\delta x)$ independent
of $k_{\rm\scriptscriptstyle F}$. The average transmitted charge, cf.\
Fig.~\ref{fig:charge}, then exhibits oscillations in the separation
$\delta x$,

\begin{align}\label{eq:res_q}
  &\langle Q/e \rangle^\text{res}_\pm  =
    2 \langle T^\text{res} \rangle \\
  &\quad \times \frac{1+ (\delta x / 2 \xi)^2 
    \pm [ \cos (k_0 \delta x) +  (\delta x / 2 \xi) \sin(k_0 \delta x) ] }
  { 1+ (\delta x / 2 \xi)^2 \pm 1}. \nonumber
\end{align}
For wave packets with a large separation $\delta x \gg \xi$, the exchange
term decays as $(\delta x)^{-2}$ and the transmitted charge is given
by $\langle Q/e \rangle^\text{res} = 2 \langle T^\text{res} \rangle$,
independent on the sign of the exchange term. On the other hand, strongly
overlapping wave packets with $\delta x \to 0$ reproduce the result
for independent particles $\langle Q/e \rangle^\text{res}_+ = 2 \langle
T^\text{res} \rangle$ for the symmetric exchange. For the anti-symmetric
case the transmitted charge
\begin{equation}\label{eq:qminus_res}
 \langle Q/e \rangle^\text{res}_- = 
 2 \langle T^\text{res} \rangle (1 - 2 \xi k_0 + 2 \xi^2 k_0^2)
\end{equation}
is reduced for narrow wave packets, $\xi k_0 < 1$ and enhanced for wave
packets with $\xi k_0>1$; the reduction can be up to $50\%$ for $\xi k_0
= 1/2$, while the increase is unlimited for $\xi k_0>1$.\cite{note} Note
that this increase is due to a large $P_{1,-}$ as $P_{2,-}$ vanishes.

The quantum point contact yields almost identical results: The off-diagonal
matrix element assumes the form $\langle 1 | T^\text{qpc} | 2 \rangle \approx
e^{-i k_0 \delta x} S \langle T^\text{qpc} \rangle$ and the \emph{interference
term} in $P_2$ vanishes.  The transmitted charge
\begin{equation}\label{eq:qpc_q}
  \langle Q/e \rangle^\text{qpc}_\pm = 
    2 \langle T^\text{qpc} \rangle
    \frac{1+ (\delta x / 2 \xi)^2 \pm \cos (k_0 \delta x) } 
    { 1+ (\delta x/ 2 \xi)^2 \pm 1} 
\end{equation}
again exhibits oscillations with $k_0 \delta x$, cf.\ Fig.~\ref{fig:charge}.
The various limits discussed above are reproduced as well, except for the
case of anti-symmetric exchange and strongly overlapping wave packets, cf.\
(\ref{eq:qminus_res}), where the average transmitted charge now is given by
\begin{equation}\label{eq:qminus_qpc}
  \langle Q/e \rangle^\text{qpc}_- 
  = 2 \langle T^\text{qpc} \rangle (1 + 2 \xi^2 k_0^2).
\end{equation}
Eqs.~(\ref{eq:qminus_res}) and (\ref{eq:qminus_qpc}) are the most striking
results of our study: for a large parameter $\xi k_0$ the mean transmitted
charge can be hugely increased as compared to the value expected for
two independent wave packets. Note that the above results are valid in a
regime where all relevant lengths remain below the phase breaking length
$L_\varphi$.  Finally, we provide the expressions for the generating
function of the full counting statistics,
\begin{align}\label{eq:2fcs}
  \chi^\text{res}_\pm =& 1 +  
  \langle Q/e \rangle^\text{res}_\pm  (e^{i\lambda} -1 )\\
  &+ \langle T^\text{res} \rangle^2 
  \frac{(1 \pm 1) [ (\delta x /2 \xi)^2 + 1]}
  { (\delta x /2 \xi)^2 + 1 \pm 1 } (e^{i \lambda} -1)^2, \nonumber\\
  \chi^\text{qpc}_\pm =& 1 + 
  \langle Q/e \rangle^\text{qpc}_\pm (e^{i\lambda} -1 )  +
   \langle T^\text{qpc} \rangle^2 (e^{i\lambda} -1 )^2.
\end{align}

The above enhancements in $\langle Q \rangle_-$ are entirely due to $P_1$;
there are two crucial elements in the game, Pauli exclusion and dispersive
scattering. Dispersive scattering generates broader wave functions and
combined with Pauli exclusion we have a reduction in $P_0$ and $P_2$, and
hence an increase in $P_1$. However, different settings may be thought of
where $P_2$ is enhanced. E.g., a large $P_{2,-}$ (see (\ref{eq:fcs_2wp})) is
obtained for wave packets with shifted amplitudes $f_2(k) = f_1(k+\delta k)$
in $k$-space and a large overlap integral $S$ combined with a transmission
amplitude suppressing $k$-values in the overlap region. A large enhancement
in the tunneling probability of two identical particles due to an enhanced
$P_2$ was also found by Suslov and Lebedev.\cite{suslov}

\subsection{Second Quantized Picture}

In a second quantized picture, electronic excitations in the quantum wire
are produced by a voltage pulse $V(t)$ applied at $x=0$.\cite{voltage}
If the time dependence of the voltage is slow compared to the transition
time of an electron through the loop, the potential can be considered as
quasistatic. It can then be incorporated in the phase factor $\exp [i \phi(t
- x/v_{\rm\scriptscriptstyle F}) \Theta(x)]$, where $v_{\rm\scriptscriptstyle
F}$ is the Fermi velocity and the phase,
\begin{equation}\label{eq:phase} 
  \phi(t) = \frac{e\Phi(t)}{\hbar c}
  = - \frac{e}{\hbar} \int_{-\infty}^t \!\! dt' V(t'),
\end{equation}
is proportional to the flux $\Phi(t)$.  The solution $\psi_{\text{L},k} (x;t)
= \exp [i k (x- v_{\rm\scriptscriptstyle F} t) + i \phi(t- x/
v_{\rm\scriptscriptstyle F}) \Theta(x) ]$ of the time dependent Schr\"odinger
equation\cite{lesovik:94} is a plane wave with well-defined energy $\hbar
v_{\rm\scriptscriptstyle F} k$, left of the position of the voltage pulse,
$x<x_\text{s}$.  The wave function $\psi_{\text{L},k}(x;t)$ for $x>0$ can be
decomposed into energy eigenmodes $\exp[ik(x-v_{\rm\scriptscriptstyle F} t) ]$
of the free Hamiltonian
\begin{equation}\label{eq:fourier}
  \psi_{\text{L},k} ( x_\text{s} > x >0;t) = 
  \! \int \!\! \frac{dk'}{2\pi} 
  U(k'-k) e^{i k' (x - v_{\rm\scriptscriptstyle F} t)}
\end{equation}
where the transformation kernel
\begin{equation}\label{eq:u} U(q) = v_{\rm\scriptscriptstyle F} \int dt e^{i \phi(t)
  + i q v_{\rm\scriptscriptstyle F} t}
\end{equation} 
is the Fourier transform of the phase factor $\exp [i \phi(t)]$.  In the
present work, we are interested in applying integer flux pulses such that
$\phi(t \rightarrow \infty) \in 2\pi \mathbb{N}$ as noninteger flux pulses do
not produce clean single-particle excitations and lead to logarithmic
divergences in the noise of the transmitted charge.\cite{lee:93,levitov:96} In
the case of integer flux pulses, the behavior of $\exp [i \phi(t)]\rightarrow
1$ for large times, $t \rightarrow \pm\infty$, leads to a Dirac delta function
$2\pi \delta (q)$.  The remaining part $U^\text{reg} (q) = U(q) - 2\pi
\delta(q)$ is finite for a localized voltage pulse;
\begin{equation}\label{eq:ureg}
  U^\text{reg} (q) = v_{\rm\scriptscriptstyle F} \int dt  \bigl[ e^{i \phi(t)} -1 \bigr]
  e^{i q v_{\rm\scriptscriptstyle F} t}.
\end{equation}
The transformation $U(k'-k)$ describes the scattering amplitude for the
transition from a momentum state $k$ (for $x<0$) to the state $k'$ (for
$x>0$) due to the application of the voltage pulse. The statement that the
wave function is in the momentum state $k'$ holds in the asymptotic region,
i.e., for observation points with $x_\text{obs} \gg \xi$.  The scattering
matrix approach at the scatterer assign the momentum component $k'$ a
transmission amplitude $\tau_{k'}$ for traversing the scatterer.  Therefore,
momentum eigenstates $k$ incoming from the left assume the form
\begin{equation}\label{eq:left}
 \psi_{\text{L},k} (x> x_\text{s} , t) =
 \int  \frac{dk'}{2\pi} 
 U(k'-k) e^{i k' (x - v_{\rm\scriptscriptstyle F} t)} \tau_{k'},
\end{equation}
right of the scattering region.  The states originating from the right do not
enter the region where the voltage pulse is applied and consists of the
incoming and the reflected part
\begin{equation}\label{eq:right}
 \psi_{\text{R},k} (x> x_\text{s} , t) =
 (e^{- i k x} + r_k e^{i k x}) e^{- i k v_{\rm\scriptscriptstyle F} t}
\end{equation}
without any shift in energy.  The time dependent field operator is given by
\begin{equation}\label{eq:field}
 \Psi (x>x_\text{s}, t) = \int \frac{dk}{2\pi} \bigl[ 
 \psi_{\text{L},k} (x;t) a_k + 
 \psi_{\text{R},k} (x;t) b_k \bigr]
\end{equation}
in the region $x>x_\text{s}$ behind the scatterer; here, $a_k$ and $b_k$
denote fermionic annihilation operators for states coming from the left, right
reservoir, respectively.  Averaging the current operator, defined as $I (x >
x_\text{s};t) = e \hbar\bigl(\Psi^\dagger(x;t) \partial_x \Psi(x;t) -
[\partial_x \Psi^\dagger(x;t)] \Psi(x;t)\bigr)/2 i m$, over the Fermi
reservoirs assuming the same Fermi distribution $n(k)$ at the far right and
left of the interaction region and integrating the ensemble averaged current
over time, the average transmitted charge
\begin{equation}\label{eq:mean_charge}
 \langle Q/e \rangle = 
  \int \frac{dk}{2\pi} n(k) \int \frac{dk'}{2\pi} K(k'-k) T_{k'}
\end{equation}
is obtained.  It depends through the kernel $K(q) = |U^\text{reg} (q)|^2 +
4\pi \delta(q) \text{Re} \bigl[ U^\text{reg} (0) \bigr]$ on different
transmission probabilities $T_{k+q}$ than incoming wave vector $k$; note that
for $U^\text{reg} (q)$ whose real part is not continuous near $q=0$,
$\text{Re} [U^\text{reg} (0)]$ has to be replaced by the symmetric limit
$\text{Re} [U^\text{reg} (0 + i\varepsilon) + U^\text{reg} (0 -
i\varepsilon)]/2$ ($\varepsilon \rightarrow 0)$.  Equivalently, the kernel
$K(q)$ can be defined as
\begin{equation}\label{eq:k}
  K(q) = v_{\rm\scriptscriptstyle F}^2 \int dt dt' 
  \bigl[ e^{i \phi(t) - i\phi(t')} -1
  \bigr] e^{i q v_{\rm\scriptscriptstyle F} (t-t')}.
\end{equation}
The first quantized result, $\langle Q/e \rangle = \int (dk/2\pi) |f(k)|^2
T_k$, is the same as the second quantized, Eq.~(\ref{eq:mean_charge}),
provided that $\int (dk'/2\pi) n(k') K(k-k') = |f(k)|^2$, where $f(k)$ denotes
the wave function in the first quantized picture.  

In a first step, we discuss a unit flux voltage pulse of Lorentzian shape
$V_{t_1}(t)= - (2 v_{\rm\scriptscriptstyle F} \xi \hbar/e)
/[v_{\rm\scriptscriptstyle F}^2( t-t_1)^2 + \xi^2]$, $v_{\rm\scriptscriptstyle
F} \xi \gg 1$ applied at time $t_1$.  The regular part of the transformation
kernel is given by
\begin{equation}\label{eq:ureg1}
  U^\text{reg}_{x_1}(q) = - 2\pi (2\xi) e^{-\xi q - i q x_1} \Theta(q),
\end{equation}
where $x_1 = v_{\rm\scriptscriptstyle F} t_1$.  The $\Theta$-function reflects
the fact that the Lorentzian pulse only increases the energy of the individual
Fourier components.  As we will discuss in Appendix~\ref{sec:equiv}, the pulse
produces a clean single particle excitation on top of the Fermi sea.
Calculating
\begin{equation}\label{eq:k1}
  K_{x_1}(q) = (2\pi)^2 2\xi \Bigl[2 \xi e^{-2 \xi q} \Theta(q) 
  - \delta(q) \Bigr]
\end{equation}
and inserting the result in Eq.~(\ref{eq:mean_charge}) yields the average
charge transmitted, (for zero temperature with $n(k) =
\Theta(k_{\rm\scriptscriptstyle F} -k)$, i.e., temperatures $\vartheta \ll
\hbar\xi/v_{\rm\scriptscriptstyle F}$)
\begin{align}\label{eq:mean_charge1}
  \langle Q/e \rangle_{x_1} &= \int_0^{k_{\rm\scriptscriptstyle F}} \!\!\!  dk \int dk'
  2 \xi \Bigl[ 2\xi e^{-2 \xi (k'-k)} \Theta(k'-k) \nonumber\\
  & \quad - \delta(k' - k) \Bigr] T_{k'} = \int \frac{dk}{2\pi}
  |f_1(k)|^2 T_k.
\end{align}
In the last step, we performed the integration over $k$ and renamed the
remaining variable of integration.  Furthermore, we dropped terms with $k
\approx 0$.  These terms correspond to the fact that the left states are
shifted up in energy and therefore, currents from filled states on the right
for small $k$ values are not canceled by corresponding currents on the left.
This is an unphysical artifact of the linear spectrum approximation which is
only valid close to the Fermi point, $k\approx k_{\rm\scriptscriptstyle F}$.
Formally, one can cure the problem by setting the lower integration bound for
the $k$ integration to $-\infty$ whenever one integrates over the Fermi sea.
The empty states are then shifted to $k \rightarrow - \infty$ and do not
appear in the final result.  Alternatively, the transmission probability $T_k$
can be set to 0 for small values of $k$.  Eq.~(\ref{eq:mean_charge1}) is
exactly the first quantized result for a wave packet of Lorentzian form,
Eq.~(\ref{eq:lorentz}).  More interesting than the single voltage pulse are
two pulses separated by $\delta x = v_{\rm\scriptscriptstyle F} \delta t =
x_2-x_1$ as they produce two wave packets and we expect interference terms to
be found as in the first quantized formalism.  For two pulses, the phase
factor is a product of the phase factors of the individual pulses.  Therefore,
the Fourier transformed function $U_{x_1, x_2}(q)$ is the convolution of the
transformation functions for single pulses $U_{x_1, x_2}(q) = U_{x_1} \!  \ast
\!  U_{x_2} (q)$,
\begin{align}\label{eq:ureg2}
  U^\text{reg}_{x_1, x_2} &(q) - U^\text{reg}_{x_1} (q)-
  U^\text{reg}_{x_2} (q) = U^\text{reg}_{x_1} \! \ast \!
  U^\text{reg}_{x_2} (q) \\
  &= \frac{2 (2\pi) (2\xi)^2}{\delta x } e^{-\xi q - i q(x_1 + x_2)/2}
  \sin(q \delta x/2) \Theta(q). \nonumber
\end{align}
In the expression for the transmitted charge, we need to know the kernel $K$
given by
\begin{multline}\label{eq:k2}
  K_{x_1, x_2}(q) = K_{x_1} (q) + K_{x_2} (q) + 2 (2\pi)^2 (2\xi)^2
  e^{-2 \xi q } \Theta(q) \\
  \times\Biggl( 2\frac{\bigl[\delta x \cos (q \delta x/2) - 2 \xi
    \sin(q \delta x/2) \bigr]^2}{(\delta x)^2} - 1 \Biggr).
\end{multline}
The first two terms correspond to the direct contribution, the last term
describes the interference.  At zero temperature the average transmitted
charge
\begin{align}\label{eq:mean_charge2}
  \langle  & Q/e \rangle_{x_1, x_2} = \langle Q/e \rangle_{x_1} 
  + \langle Q/e \rangle_{x_2}   \\
  & + 2 \int_0^\infty \!\!\!\! dq \,2 \xi e^{-2 \xi q} T_{k_{\rm\scriptscriptstyle F} + q}
  \frac{1 - \cos(q \delta x) - \frac{\delta x}{2\xi}
    \sin( q \delta x) }{(\delta x /2 \xi)^2} \nonumber\\
  & = P_{1,-} + 2 P_{2,-}, \nonumber
\end{align}
for two Lorentzian voltage pulses separated by $\delta x$ is obtained.  When
comparing Eq.~(\ref{eq:mean_charge2}) to the first quantized result
(\ref{eq:prob}), it can be seen that the result in second quantization is the
same as expected from the analysis in the first quantized picture for an
antisymmetric wave function.  For a single voltage pulse with double flux,
i.e.  $\delta x=0$, the increase in the transmitted charge as seen in the
first quantized formalism is confirmed.  Moreover, applying a voltage pulse
carrying even more flux quantum, leads to a further increase in the
transmitted charge with respect to the independent particle picture, cf.\
Appendix~\ref{sec:many_flux}.

So far, we considered only spinless particles. The electronic spin $1/2$
is included easily in the second quantized formalism.  The summation over
the spin degree of freedom provides a factor of two for each cumulant;
i.e., each voltage pulse excites two particles in the same spatial state.
In the first quantized formalism, a single voltage pulse which excites
two particles in the state $\Psi_1 (x;t)$ can be described as a singlet
\begin{equation}\label{eq:spin1wf}
  \Psi (x_1, x_2, t) = \Psi_1(x_1 , t) \Psi_1 (x_2, t) \frac{| {\uparrow
      \downarrow} \rangle - | {\downarrow \uparrow} \rangle }
  {\sqrt{2}}.
\end{equation}
Each of the two particles carries the same current and they are independent of
each other which leads to a factor two as in the second quantized formalism.
Accordingly, two unit flux voltage pulses excite a four particle state
involving the wave functions $\Psi_{1,2} (x;t)$.  The second quantized
formalism leads a factor of two due to the spin degeneracy and an averaging
over the energies of the antisymmetrized wave function $\Psi_- (x_1, x_2;t) =
\bigl[ \Psi_1 (x_1;t) \Psi_2(x_2, t) - (x_1 \leftrightarrow x_2) ] \bigr] /
\sqrt{2(1 - |S|^2) }$.  In first quantization, the four-particle wave function
with spin 0 is given by
\begin{align}\label{eq:spin2wf}
 \Psi (x_1, x_2&, x_3, x_4, t) = \\
 \frac{1}{\smash[b]{\sqrt{6}}} \Bigl[
 &\Psi_- (x_1, x_2, t) \Psi_- (x_3, x_4, t) \bigl( 
 | {\uparrow\uparrow\downarrow\downarrow}\rangle 
 + | {\downarrow\downarrow\uparrow\uparrow}\rangle \bigr) \nonumber\\
 + &\Psi_- (x_1, x_4, t) \Psi_- (x_2, x_3, t)\bigl( 
 | {\uparrow\downarrow\downarrow\uparrow}\rangle 
 + | {\downarrow\uparrow\uparrow\downarrow}\rangle \bigr) \nonumber\\
 - &\Psi_- (x_1, x_3, t) \Psi_- (x_2, x_4, t)\bigl( 
 | {\uparrow\downarrow\uparrow\downarrow}\rangle 
 + | {\downarrow\uparrow\downarrow\uparrow}\rangle \bigr) \Bigr]. \nonumber
\end{align}
As the spin is unimportant, we may integrate over the spin degrees of freedom
and obtain a density matrix consisting of an equal mixture of the wave
function $\Psi_- (x_1, x_2;t) \Psi_- (x_3, x_4;t)$ and wave functions obtained
by coordinate relabeling.  Relabeling does not change the physics of
independent particles.  We can choose any one of the wave functions in the
first quantized picture.  Both the antisymmetry ($\Psi_-$) and the trivial
factor of two (product of two wave functions) are now also visible in the first
quantized language.  The generating function $\chi_\pm^{(4)} = \chi_\pm{}^2$
of the full counting statistics for the four spin degenerate particles is the
square of the spinless generating functions $\chi_\pm$ of Eq.~(\ref{eq:2fcs}).

The procedure outline can also be carried out for different forms of the
voltage pulse.  Note though that for pulses other than a Lorentzian voltage
pulse there is no simple first quantized correspondence to the second
quantized formalism.  A unit flux voltage pulse is given by
\begin{equation}\label{eq:2nd_pulse}
 V(t) = - \frac{2 \hbar \dot{f}(t)/e}{1 + f(t)^2}
\end{equation}
with the requirement that $f(t\rightarrow \pm \infty ) \rightarrow \pm
\infty$; the Lorentzian voltage pulse, considered so far, is a special case
obtained with $f(t) = v_{\rm\scriptscriptstyle F} t/ \xi$.  The phase factor
assumes the form
\begin{equation}\label{eq:2nd_phase}
 e^{i \phi(t)} = \frac{f(t) - i}{f(t) + i}.
\end{equation}
As an example we consider the voltage pulse given by $f(t) =
(v_{\rm\scriptscriptstyle F} t / \xi)^3$ which has a simple transformation
function $U(q) = 2\pi \delta(q) + U^\text{reg} (q)$ with
\begin{align}\label{eq:2nd_ureg}
  U^\text{reg}(q) &=- 2\pi \frac{2 \xi}{3} \bigl[
  e^{\xi q} \Theta(-q) \\
  &+ e^{-\xi q/2} [ \cos (\sqrt{3} \xi q/2) + \sqrt{3} \sin(\sqrt{3} \xi
  q/ 2) ] \Theta(q) \bigr] \nonumber
\end{align}
The part with positive momentum transfer, $q>0$, corresponds to an excited
electron as we have seen for the simple Lorentzian pulse. The negative
momentum indicates holes accompanying the electron. The kernel $K$ assumes
the form
\begin{align}\label{eq:2nd_kernel}
  K(q)&= (2 \pi )^2 \frac{(2\xi)^2}{9} \bigl[ e^{2\xi q} \Theta(-q) \\
  & + e^{-\xi q} [2 - \cos( \sqrt{3} \xi q ) 
  + \sqrt{3} \sin( \sqrt{3} \xi q)] \Theta(q) \bigr] \nonumber \\
  & - 2 (2 \pi)^2 (2 \xi) \delta(q)/ 3. \nonumber
\end{align}
Plugging it in Eq.~(\ref{eq:mean_charge}), yields the average
transmitted charge
\begin{align}\label{eq:2nd_current}
  \langle Q/e \rangle =& \frac{2 \xi}{9} \int dk \, T_k \Bigl[
  - e^{- 2\xi(k_{\rm\scriptscriptstyle F} - k)} \Theta( k_{\rm\scriptscriptstyle F} - k ) \nonumber\\
  &+ e^{-\xi (k - k_{\rm\scriptscriptstyle F})} \Bigl( 4 + \cos \bigl[\sqrt{3}
  \xi ( k - k_{\rm\scriptscriptstyle F})\bigr]
  \nonumber\\
  &+ \sqrt{3} \sin \bigl[\sqrt{3} \xi (k - k_{\rm\scriptscriptstyle F}) \bigr]
   \Bigr)\Theta(k-k_{\rm\scriptscriptstyle F})  \Bigr]
\end{align}
at zero temperature.  The part with the negative momentum transfer carries a
negative charge, hole, and the charge of the other part is positive and so the
expected nature of excitations is confirmed.  Note that in the case of a fully
transparent wire ($T_k \equiv 1$), the average transmitted charge is exactly
one electron $e$.

\section{Multi-Lead}

In a multi-lead setup, exchange effects can be found for an energy-independent
scattering already.  Here, we discuss the three-lead fork geometry and the
four-lead reflectionless beam splitter for two incoming particles in different
leads 1 and 2 and hence vanishing initial overlap.  The wave packets are
assumed to have a Lorentzian shape (\ref{eq:lorentz}) and to impinge on the
scatterer with a delay $\delta t = \delta x/v_{\rm\scriptscriptstyle F}$.  In
the three-lead geometry, cf.\ Fig.~\ref{fig:setup}(b), the generating function
reads
\begin{equation}\label{eq:gen3}
  \chi^{\threelead}_\pm = 
  1 + T_3 (e^{i \lambda} -1) 
  + (1 \pm |S|^2) T_{13} T_{23} (e^{i \lambda} -1)^2,
\end{equation}
where $T_{13}$ and $T_{23}$ denote the transmission probabilities for
particles incident from leads $1$ and $2$ and propagating into lead $3$
(containing the counter) and $T_{3}=T_{13}+T_{23}$. The probabilities
\begin{align}\label{eq:prob3}
  P^{\threelead}_{0,\pm} &= (1 - T_{13}) (1-T_{23}) 
  \pm |S|^2 T_{13} T_{23} \nonumber\\
  P^{\threelead}_{1,\pm} &= T_{13} ( 1 - T_{23}) + (1-T_{13}) T_{23} 
  \mp 2 |S|^2 T_{13} T_{23} \nonumber\\
  P^{\threelead}_{2,\pm} &= (1 \pm |S|^2 ) T_{13} T_{23}
\end{align}
depend on the exchange symmetry as long as $T_{13} T_{23} \ne 0$. For
$\delta x=0$ the overlap integral $S = e^{-i k_{\rm \scriptscriptstyle
F} \delta x}/(1+i\delta x/2\xi)$ becomes unity and hence $P_{2,-}=0$.
Finally, the charge cumulants assume the form
\begin{align}\label{eq:cum3}
  \langle Q/e \rangle^{\threelead}_\pm \!&=\! T_3, \\
  \langle\langle (Q/e)^2 \rangle\rangle^{\threelead}_\pm \! &= \!
  T_{13}(1\!-\!T_{13}) + 
  T_{23} (1\!-\!T_{23}) \pm \! 2 |S|^2 T_{13} T_{23}; \nonumber
\end{align}
effects due to exchange symmetry are limited to the charge noise which is
enhanced (reduced) for the (anti-)\linebreak symmetric case.  For the
reflectionless four-lead beam splitter with the counter in lead 3,
Fig.~\ref{fig:setup}(c), we obtain the generating function, probabilities, and
moments in the form
\begin{align}\label{eq:gen4}
  \chi^{\fourlead}_\pm &= 1 + [ T_3 \pm |S|^2 ( T_3 -1 ) ] 
  (e^{i \lambda} -1) \nonumber \\
  & \quad + (1 \pm |S|^2)  T_{13} T_{23} (e^{i \lambda} -1 )^2 \nonumber\\
  P^{\fourlead}_{0,\pm} &= (1 \pm |S|^2) (1 - T_{13}) (1 - T_{23}) \nonumber\\
  P^{\fourlead}_{1,\pm} &= (1 \pm |S|^2) [ (1 - T_{13}) T_{23}
  + T_{13} (1- T_{23}) ] \nonumber\\
  & \quad\mp |S|^2 \nonumber\\
  P^{\fourlead}_{2,\pm} &= (1 \pm |S|^2) T_{13} T_{23} \nonumber\\
  \langle Q/e \rangle^{\fourlead}_\pm &= ( 1 \pm |S|^2) T_3
  \mp |S|^2 \nonumber\\
  \langle\langle (Q/e)^2 \rangle\rangle^{\fourlead}_\pm & = (1 \pm
  |S|^2) \bigl[
  (1 - T_{13}) T_{13} + (1 - T_{23}) T_{23} \nonumber\\
  & \quad \mp |S|^2(T_3 -1 )^2 \bigr].
\end{align}
Particles with anti-symmetric exchange and $\delta x=0$ exhibit
no partitioning and the noise vanishes,\cite{burkard:00} as
$P^{\fourlead}_{0,-} = P^{\fourlead}_{2,-} = 0$ and $P^{\fourlead}_{1,-}
= 1$, i.e., each of the incoming particles is transmitted in a different
outgoing lead.\cite{burkard:00} In the nonsymmetric case $T_3\ne 1$ (note
that $T_{13}+T_{14}=1$ and $T_{14}=T_{23}$ in the symmetric situation),
an exchange effect shows up already in the average transmitted charge
$\langle Q/e \rangle^{\fourlead}_\pm=(1\pm |S|^2)T_3 \mp |S|^2$.

\section{Experimental Verification}

In order to observe exchange effects in an experiment, we propose to test for
the predicted non-linearity in the transport: comparing the average
transmitted charge $\langle Q^{\scriptscriptstyle (1)}\rangle$ for a single
flux pulse with the one ($\langle Q^{\scriptscriptstyle (2)}\rangle$) injected
with doubled voltage, our analysis predicts the non-trivial result $\langle
Q^{\scriptscriptstyle (2)}\rangle/2\langle Q^{\scriptscriptstyle (1)}\rangle =
1+2\xi^2 k_0^2$, cf., Eqs.\ (\ref{eq:qminus_qpc}); for a transmission
resonance, $\langle Q^{\scriptscriptstyle (2)}\rangle/2\langle
Q^{\scriptscriptstyle (1)}\rangle = 1-2\xi k_0+2\xi^2 k_0^2$, cf., Eqs.\
(\ref{eq:qminus_res}). Given a pulse length of duration $\sim 10^{-10}~{\rm
s}$, such an experiment requires a device with a phase coherence length beyond
$10~\mu{\rm m}$ and is preferably carried out on a quantum point contact where
Coulomb effects are less prominent. In this case, we have to compare the
spread in energy $\hbar v_{\rm \scriptscriptstyle F}/\xi$ of the wave packet
with the charging energy $e^2/C$, where $C \sim \varepsilon L$ denotes the
junction capacitance ($L$ the length of the point contact).  For a sharp
transmission step, we have $\xi/L<1$ small and the relation $\xi/L <
\varepsilon (\hbar c/e^2)(v_{\rm \scriptscriptstyle F}/c)$ can be satisfied.

\section{Conclusion} 

In summary, we have studied effects of the exchange symmetry on the
transport statistics in mesoscopic systems.  We find a strong dependence
(and possibly huge enhancement) of the average charge transmission on the
exchange symmetry, provided that the wave function sufficiently overlap. For
antisymmetric exchange such overlap pushes weight of the two-particle wave
function to higher energies and combined with the energy dependent scattering
this leads to the observed enhancement. For multi-channel/multi-lead
setups, exchange effects can already be observed for energy-independent
scattering probabilites. There the origin of the effect is rooted in simple
Pauli-blocking. Furthermore, we have proposed a experiment to test for
the most striking effect of exchange; the non-linearity in transport.

We thank M.\ Indergand, A.V.\ Lebedev, M.V.\ Lebedev, and M.V.\ Suslov
for discussions and acknowledge financial support from the CTS-ETHZ and
the Russian Foundation for Basic Research (06-02-17086-a).

\appendix
\bigskip

\section{Equivalence Between the First and Second Quantized
  Picture}\label{sec:equiv}

Finally, we motivate the form (\ref{eq:lorentz}) for the $k$-space amplitude
$f_1(k)$ of the pulse-generated wave packet.  In the adiabatic limit, a
voltage pulse $V(t)$ applied at $x=0$ adds a phase $\phi(t)=-(e/\hbar)
\int_{-\infty}^t dt' V(t')$, $e>0$, to the wave function across the origin
\cite{lesovik:94}.  We derive the scattering matrix $U_{k',k}$ describing the
transitions from $k$ states at $x<0$ to $k'$ states at $x>0$.  This is
conveniently done in real space, where a particle described by the
time-retarded state $|\tilde{x} \rangle$ with $\langle x,t|\tilde{x} \rangle =
\delta(x- v_\text{F}t)$ for $x<0$ evolves to $\exp [i \phi(-\tilde{x}/
v_\text{F})]|\tilde{x}\rangle$ at $x>0$ under the action of the voltage pulse
($\tilde{x}=x-v_{\rm \scriptscriptstyle F} t$ is the retarded variable).  An
arbitrary incoming state at $x<0$ then is transformed into the corresponding
outgoing state at $x>0$ via the unitary scattering operator $\hat
U=\exp[i\int\!  d \tilde{x} \phi(-\tilde{x}/v_{\rm \scriptscriptstyle F})
\Psi^\dagger (\tilde{x}) \Psi(\tilde{x})]$ with the time-retarded field
operator $\Psi(\tilde x)$.  We introduce the fermionic annihilation operators
$a_k$, generating basis states $\exp (i k \tilde{x})$, and transform to
$k$-space via $\Psi(\tilde x) = \int (dk/2\pi)\exp (i k \tilde x) a_k$ to find
the matrix elements $U_{k',k} = U (k'-k) = v_{\rm \scriptscriptstyle
F}\!\int\!dt \exp[i\phi(t)+i(k'-k) v_{\rm \scriptscriptstyle F} t]$.  The
scattering amplitude then relates to the amplitude $f_1(k)$ in
(\ref{eq:lorentz}) via $U_{x_1}(q)\!=\!-\sqrt{4\pi \xi}f_1(k_{\rm
\scriptscriptstyle F} + q)$ $e^{ik_{\rm \scriptscriptstyle F}v_{\rm
\scriptscriptstyle F} t_1}+2\pi\delta(q)$, cf.\ Eq.~(\ref{eq:ureg1}).  The
voltage pulse transforms the Fermi sea $|{\Phi}_{\rm \scriptscriptstyle
F}\rangle$ to $|\bar{\Phi}_{\rm \scriptscriptstyle F}\rangle = \hat{U}_{x_1}
|{\Phi}_{\rm \scriptscriptstyle F}\rangle$; the calculation of the density
matrix at zero temperature
\begin{align}\label{eq:matrix1} 
  \langle \bar{\Phi}_{\rm\scriptscriptstyle F}|
  a^\dagger_{k'} & 
  a^{\vphantom{\dagger}}_k | \bar{\Phi}_{\rm\scriptscriptstyle F} \rangle 
  \! =\!\!\!  \int \!\!  \frac{dk''}{2\pi} U_{x_1}^*\! (k' - k'') 
  U_{x_1} \!(k - k'')
 \Theta(k_{\rm\scriptscriptstyle F}-k'')\nonumber\\ =&\langle
 {\Phi}_{\rm\scriptscriptstyle F} | a^\dagger_{k'} a^{\vphantom{\dagger}}_k
 |{\Phi}_{\rm\scriptscriptstyle F}\rangle + f_1^*(k') f^{\vphantom{*}}_1(k),
\end{align}
involves the transformed annihilation operators
\begin{equation}\label{eq:trans}
  \hat U^\dagger a_k \hat U = 
  a_k + \int \frac{dk'}{2\pi} U^\text{reg} (k - k') a_{k'}
\end{equation}
with $U^\text{reg} (q)$ defined in Eq.~(\ref{eq:ureg}).  It tells us that the
voltage pulse leaves back a filled Fermi sea above which an excitation is
created with an amplitude $f_1(k)$ given by Eq.\ (\ref{eq:lorentz}).
Furthermore, we find that the two-particle correlator $\langle \bar{\Phi}_{\rm
\scriptscriptstyle F}| a^\dagger_{k} a^\dagger_{k'}
a^{\vphantom{\dagger}}_{k'} a^{\vphantom{\dagger}}_k |\bar{\Phi}_{\rm
\scriptscriptstyle F} \rangle$ vanishes for $k,k' > k_{\rm\scriptscriptstyle
F}$ and hence pair-excitations are absent.  To observe the effects of the
exchange symmetry, the overlap integrals are important.  To create two
particles, two pulses need to be applied in sequence, $\hat U= \hat U_{x_2}
\hat U_{x_1}$.  A straightforward calculation shows that
\begin{widetext} 
  \begin{equation}\label{eq:matrix2}
  \langle \Phi_{\rm\scriptscriptstyle F} | \hat U_{x_1}{\!}^\dagger \hat U_{x_2}{\!}^\dagger
  a^\dagger_{k'} a^{\vphantom{\dagger}}_k \hat U_{x_2} \hat
  U_{x_1} | \Phi_{\rm\scriptscriptstyle F} \rangle = \langle \Phi_{\rm\scriptscriptstyle F} | a^\dagger_{k'}
  a^{\vphantom{\dagger}}_k | \Phi_{\rm\scriptscriptstyle F}\rangle +  \int \frac{dk''}{2\pi}
 \frac{[f^*_1(k') f^*_2(k'') - f^*_2(k') f^*_1(k'')] [f_1(k) f_2(k'')
    - f_2(k) f_1(k'')]}{1 - |S|^2}, \nonumber
\end{equation} 
\end{widetext} 
i.e., the resulting state is the unperturbed Fermi system with an
antisymmetrized two particle wave packet on top of it.  Therefore, the first
quantized picture can be used even for two voltage pulses resulting in a two
particle wave function in the first quantized picture and
Eq.~(\ref{eq:fcs_2wp}) gives the full counting statistics of the two
transmitted electrons.

\section{Many Flux Lorentzian Voltage Pulse}\label{sec:many_flux}

Applying a Lorentzian shaped $n$ flux voltage pulse $V_n (t) = - (2 n
v_{\rm\scriptscriptstyle F} \xi \hbar/e)/[(v_{\rm\scriptscriptstyle F} t)^2 + \xi^2]$, an even more drastic
nonlinear effect is expected than found for the two particle case. The
phase factor $\exp [i \phi(t)]$ is the $n$-th power of the single pulse
phase factor. The Fourier transformation of the phase factor can be carried
out using Cauchy's formula. We obtain
\begin{equation}\label{eq:u_n}
  U^\text{reg}_n (q) = 2\pi (2\xi) e^{-\xi q} \Theta(q) 
  \sum_{l=1}^n \frac{n! (-1)^l (2 \xi q)^{l-1} }{(l-1)! l! (n-l)!}.
\end{equation}
The kernel is then given by
\begin{multline}\label{eq:k_n}
  K_n (q) = - (2\pi)^2 n (2\xi) \delta(q) + (2\pi)^2 (2\xi)^2 e^{-2 \xi q}
  \Theta(q) \\ \times\sum_{l,m=1}^n \frac{ (n!)^2 (-1)^{l+m} (2\xi
    q)^{l+m-2}}{(l-1)! l! (n-l)! (m-1)! m!  (n-m)!}
\end{multline} 
To calculate the average transmitted charge, Eq.~(\ref{eq:k_n}) need to
be integrated over the Fermi sea, c.f.\ Eq.~(\ref{eq:mean_charge}). The
averaging over the zero temperature Fermi sea leads to
\begin{multline}\label{eq:kav_n}
  |f(k)|_n^2 = \int \frac{dk'}{2\pi} n(k') K(k-k') = 2\pi (2\xi)
  \Theta(k-k_{\rm\scriptscriptstyle F}) \\
  \times\sum_{l,m=1}^n \frac{ (n!)^2 (-1)^{l+m} \Gamma(l+m-1, 2 \xi
  (k- k_{\rm\scriptscriptstyle F}) )} {(l-1)! l! (n-l)! (m-1)! m!  (n-m)!}
\end{multline}
where $\Gamma(n,x_0) =\int_{x_0}^\infty dx\, x^{n-1} e^{-x}$ denotes the
incomplete gamma function. The average charge transmitted is given by
$\langle Q/e \rangle_n = \int (dk/2\pi) |f(k)|_n^2 T_k$. The result for
the scattering resonance can then be calculated,
\begin{align}\label{eq:res_n}
 \langle Q/e \rangle_2 &= 2 \langle T^\text{res}\rangle (1 - 2 \xi
 k_0 + 2 \xi^2 k_0^2) \nonumber\\
 \langle Q/e \rangle_3 &= 3 \langle T^\text{res}\rangle (1 - 4 \xi k_0
 + 8 \xi^2 k_0^2 - 16 \xi^3 k_0^3/3 \nonumber\\
 &\quad+ 4 \xi^4 k_0^4/3) \nonumber\\
 \langle Q/e \rangle_4 &= 4 \langle T^\text{res}\rangle (1 -6 \xi k_0
 + 18 \xi^2 k_0^2 -68 \xi^3 k_0^3/2 \nonumber \\
 &\quad+ 14 \xi^4 k_0^4 - 4 \xi^5 k_0^5 +
 4 \xi^6 k_0^6/9) ,
\end{align}
using the fact that the incomplete gamma function for integer $n$ is
expressible through a polynomial times an exponential. For the quantum
point contact, the equivalent expressions read
\begin{align}\label{eq:qpc_n}
  \langle Q/e \rangle_2 &= 2 \langle T^\text{qpc}\rangle (1 + 2 \xi^2
 k_0^2) \nonumber\\
  \langle Q/e \rangle_3 &= 3 \langle T^\text{qpc}\rangle (1 + 4 \xi^2
  k_0^2 - 8 \xi^3 k_0^3/3 + 4 \xi^4 k_0^4/3) \nonumber\\
 \langle Q/e \rangle_4 &= 4 \langle T^\text{qpc}\rangle (1 + 6 \xi^2
 k_0^2 - 8 \xi^3 k_0^3 + 22 \xi^4 k_0^4/3 \nonumber \\
 & \quad - 8 \xi^5 k_0^5/3 + 4 \xi^6 k_0^6/9).
\end{align}
Increasing the voltage, $V_n \propto n$, which corresponds to sending
more electrons through the scattering region, there is a huge nonlinear
increase in the transmitted charge if $\xi k_0 \gg 1$.


\begin{thebibliography}{10}
\bibitem{note}
The exponential decrease $\propto \exp(-2k_0\xi)$ in $\langle
  T^\text{res}\rangle$ guarantees that $\langle Q/e \rangle^\text{res}\ll 2$
  for $\xi k_0 >1$.
\bibitem{voltage}
Note that in the experiment \cite{kozhevnikov:00} on photon-assisted noise
  \cite{lesovik:94} instead of a voltage pulse localized at a specific position
  a bias voltage over the whole sample was used. This procedure might not work
  in our case as our effect is more sensitive and depends on the details of the
  voltage drop across the sample.
\end{thebibliography}
\end{document}